\documentclass{preprint}
\setcitestyle{square,numbers,comma}

\usepackage{amssymb}
\usepackage{amsmath}
\usepackage{lipsum}
\usepackage{booktabs}
\usepackage{makecell}
\usepackage{setspace}
\usepackage{bibentry}
\definecolor{Pink}{RGB}{255,105,180}
\usepackage{hyperref}       
\hypersetup{
    colorlinks = true,
    linkcolor=Pink,
    citecolor=Pink,
    urlcolor=Pink
}

\newcommand{\myabstract}{Advances in deep learning have opened an era of abundant and accurate predicted protein structures; however, similar progress in protein ensembles has remained elusive. This review highlights several recent research directions towards AI-based predictions of protein ensembles, including coarse-grained force fields, generative models, multiple sequence alignment perturbation methods, and modeling of ensemble descriptors. An emphasis is placed on realistic assessments of the technological maturity of current methods, the strengths and weaknesses of broad families of techniques, and promising machine learning frameworks at an early stage of development. We advocate for ``closing the loop" between model training, simulation, and inference to overcome challenges in training data availability and to enable the next generation of models.}

\newcommand{\mytitle}{AI-based Methods for Simulating, Sampling,\\ and Predicting Protein Ensembles}

\begin{document}

\title{\mytitle}
\abstract{\myabstract}
\author[1]{Bowen Jing}
\author[1,2]{Bonnie Berger}
\author[1]{Tommi Jaakkola}
\affiliation[1]{CSAIL, Massachusetts Institute of Technology}
\affiliation[2]{Dept. of Mathematics, Massachusetts Institute of Technology}
\correspondence{\email{bjing@mit.edu}, \email{bab@mit.edu}, \email{tommi@csail.mit.edu}}
\maketitle

\section{Introduction}
Sampling the equilibrium ensemble of proteins and other biomolecules is a significant and longstanding problem in structural biology \cite{karplus2002molecular, hollingsworth2018molecular}. Such ensembles can provide fine-grained functional characterization or mechanistic understanding at levels difficult to directly interrogate with experiments. Recent AI-based techniques \cite{jumper2021highly,abramson2024accurate,boitreaud2024chai,wohlwend2024boltz} have made dramatic advances in accurately predicting protein structures, but these represent static, incomplete snapshots of dynamic conformational ensembles. Thus, accurate, cheap computational methods for obtaining protein ensembles would significantly expand our scientific toolbox and accelerate computation--experiment iteration loops in hypothesis generation and design.

Molecular dynamics (MD), when targeting the Boltzmann distribution via thermostatted equations of motion, has long been the traditional workhorse for computational emulation of biomolecular ensembles \cite{hollingsworth2018molecular}. Despite its remarkable versatility, MD struggles with a vast separation of timescales between femtosecond-level integration steps and millisecond-level transitions often required for full exploration of the conformational landscape. Enhanced sampling techniques have partially bridged this gap \cite{sugita1999replica,laio2002escaping}, but are sensitive to hyperparameters and difficult to apply in the general setting. Even $\mu$s-scale simulations---which can reveal nontrivial motions but usually do not approach convergence---require multiple GPU-days, precluding any high-throughput applications. This stands in stark contrast to structure prediction models, which can yield meaningful predictions in GPU-seconds or -minutes.

\begin{figure}
    \centering
    \includegraphics[width=\linewidth]{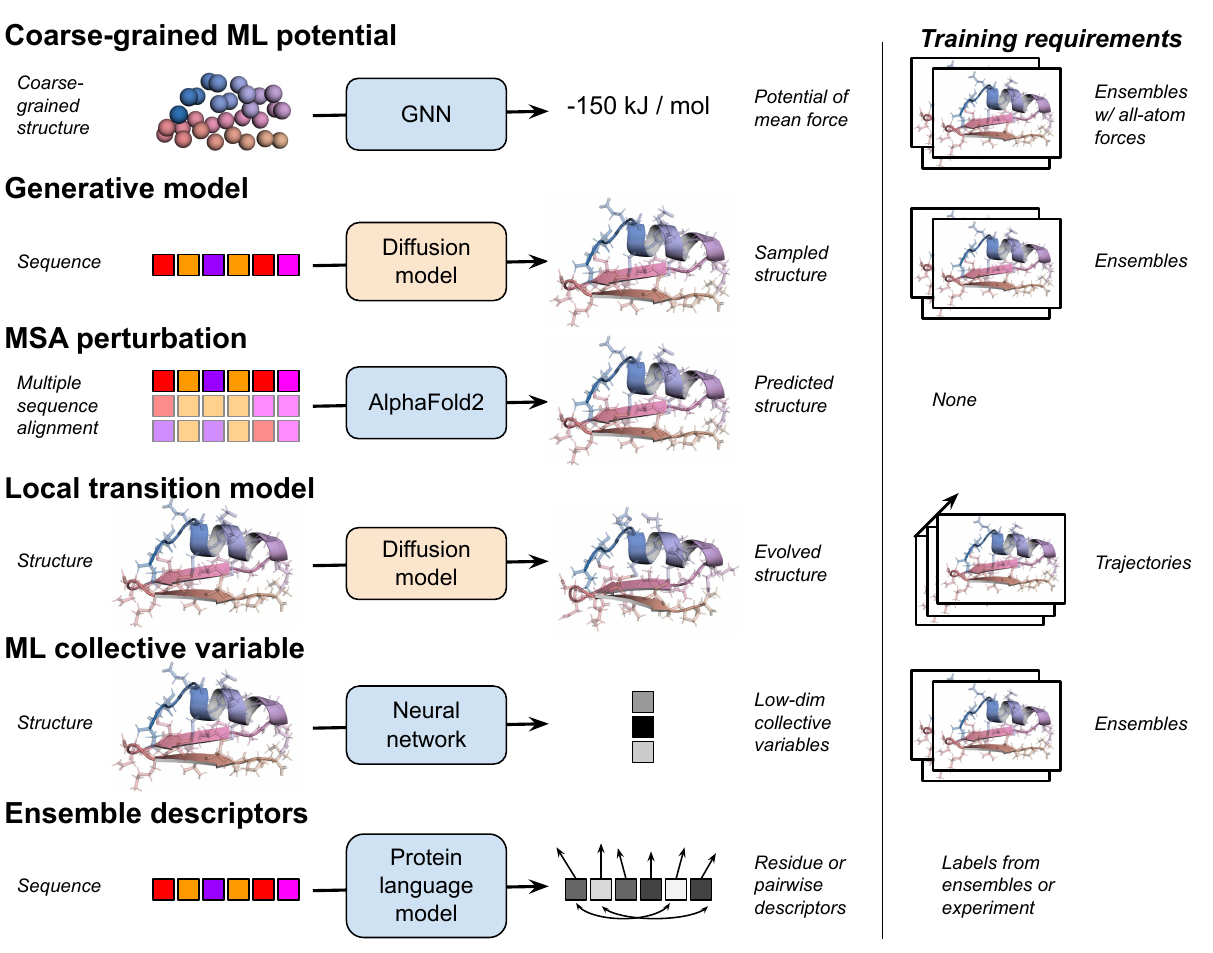}
    \caption{Classification of AI-based methods for protein ensembles according to the functions being approximated or learned by the neural network. Labels describe representative inputs, model architectures, outputs, and training requirements of each family of methods; see main text for details. Models shaded with light orange are generative models and therefore also implicitly take in noise as input. For transferable methods, the network also takes in sequence as input.}
    \label{fig:overview}
\end{figure}

In this review, we outline recent AI-based methods for simulating, sampling, or predicting conformational ensembles of proteins, using only sequence or a single structure as input. In lieu of brute-force numerical simulation with highly rugged physical potentials, these methods seek to leverage a variety of data-driven strategies to accelerate the exploration of the conformational landscape. As the defining property of modern AI is the use of deep neural networks as expressive function approximators, our review organizes methods according to \textit{what functions they seek to approximate} (Figure~\ref{fig:overview}). We will assess the technological maturity of these methods according to (1) the accuracy of the predicted ensembles; (2) the system size for which the method has been developed and demonstrated; and (3) the transferability to systems outside the training set. However, we will also highlight conceptually significant, emerging machine learning frameworks even if they have yet to approach realistic use cases. We will offer perspectives on how to best combine existing and emerging strategies towards an accurate, scalable, and transferable emulator of protein ensembles. 

\begin{table}
    \caption{AI-based methods for protein ensembles. Methods are labeled with the largest system with validations from the original publication. ``(PDB)" denotes that the publication showed results on a large number of proteins with sequence lengths representative of the Protein Data Bank (PDB). Total simulation time is reported for training data. Times may differ for methods using the same datasets due to different splits. $^1$Additional sources of training data are not listed; see original publication. $^2$AF2RAVE is transferable in the sense that the protocol can be applied to any protein, although the collective variable itself is not.}
    \label{tab:overview}
    \centering
    \begin{small}
    \begin{tabular}{lccc}
    \toprule
    Method & Largest system & Transferability & Training data \\
    \midrule
    \multicolumn{3}{l}{\textbf{Coarse-grained ML potentials}} \\ 
    Majewski et al \cite{majewski2023machine} & 80 AA & 12 fast-folders & 9 ms MD\\
    Charron et al \cite{charron2025navigating} & 189 AA & Monomers+PPIs & 100 $\mu$s MD \\
    \midrule
    \multicolumn{3}{l}{\textbf{Generative models}} \\ 
    DiG \cite{zheng2024predicting} & 306 AA & Monomers & PDB+100 $\mu$s MD+force field\\
    AlphaFlow \cite{jing2024alphafold} & (PDB) & Monomers & PDB+380 $\mu$s MD \cite{vander2024atlas} \\
    UFConf \cite{fan2024accurate} & (PDB) & Monomers & PDB \\
    BioEmu \cite{lewis2025scalable} & (PDB) & Monomers & AFDB+200 ms MD$^1$ \\
    aSAM \cite{janson2025deep} & (PDB) & Monomers & 350 $\mu$s MD \cite{vander2024atlas}+62 ms MD \cite{mirarchi2024mdcath} \\
    \midrule
    \multicolumn{3}{l}{\textbf{Generative models (exact likelihood)}} \\ 
    Boltzmann gen. \cite{noe2019boltzmann} & BPTI (58 AA) & No & 1 ms MD \cite{shaw2010atomic}+force field\\ 
    Transferable BG \cite{klein2024transferable} & 2 AA & Dipeptides & 10 $\mu$s MD \cite{klein2023timewarp}\\ 
    Prose \cite{tan2025amortized} & 8 AA & Peptides & 771 $\mu$s MD \\
    \midrule
    \multicolumn{3}{l}{\textbf{Generative models (from energy)}} \\ 
    FAB \cite{midgley2023flow} & alanine dipeptide & No & force field \\
    TA-BG \cite{schopmans2025temperature} & alanine hexapeptide & No & force field\\
    PITA \cite{akhound2025progressive} & alanine tripeptide & No & force field \\
    \midrule
    \multicolumn{3}{l}{\textbf{MSA perturbation}} \\ 
    MSA subsampling \cite{del2022sampling} & (PDB) & Monomers & None\\
    AFCluster \cite{wayment2024predicting} & (PDB) & Monomers & None \\
    \midrule
    \multicolumn{3}{l}{\textbf{Local transitions}} \\ 
    TimeWarp \cite{klein2023timewarp} & 4 AA & Peptides & 80 $\mu$s MD \\
    ITO \cite{schreiner2023implicit} & 35 AA & 4 fast-folders & 977 $\mu$s MD \cite{lindorff2011fast} \\
    EquiJump \cite{costa2024equijump} & 80 AA & 12 fast-folders & 8.2 ms MD \cite{lindorff2011fast} \\
    Str2Str \cite{lu2024strstr} & 80 AA & Monomers & PDB \\
    \midrule
    \multicolumn{3}{l}{\textbf{ML collective variables}} \\ 
    AF2RAVE \cite{vani2023alphafold2} & 369 AA & Yes$^2$ & MD, length varies \\
    \midrule
    \multicolumn{3}{l}{\textbf{Ensemble descriptors}} \\ 
    SeqDance \cite{hou2024seqdance} & (PDB) & Monomers & 450 $\mu$s MD  \cite{vander2024atlas}$^1$\\
    SeaMoon \cite{lombard2025seamoon} & (PDB) & Monomers & PDB \\
    Dyna-1 \cite{wayment2025learning} & (PDB) & Monomers & NMR \\
    RocketSHP \citep{sledzieski2025rocketshp} & (PDB) & Monomers & 417 $\mu$s MD \cite{vander2024atlas}+62 ms MD \cite{mirarchi2024mdcath} \\
    DynaProt \citep{bafna2025learning}  & (PDB) & Monomers & 380 $\mu$s MD \cite{vander2024atlas} \\
    \bottomrule
    \end{tabular}
    \end{small}
\end{table}

\section{Coarse-grained ML potentials}

A significant factor in the difficulty of simulating ensembles via MD is the high dimensionality of the potential energy surface. Hence, \textit{coarse-grained potentials} have been long sought to reduce the number of variables in the force field and provide a smoother potential energy surface. Traditional coarse-grained potentials grouped atoms into beads and modeled interactions between them using functional forms similar to classical all-atom potentials \cite{souza2021martini}. However, these force fields have generally been too inaccurate to replicate the behavior observed in all-atom simulation, often requiring restraints even to maintain the native protein fold \cite{poma2017combining}.

Machine learning (ML) potentials that parameterize the energy surface with expressive neural networks have presented an alternative paradigm for constructing coarse-grained force fields. Since the training target of the neural network, the so-called \textit{potential of mean force} (PMF), is not available in closed form, a number of frameworks have been developed to train coarse-grained force fields using only the underlying all-atom potential. The most mature and widely-used technique is \textit{variational force matching}, in which the coarse-grained force is trained to match the conditional expectation of the all-atom forces \cite{noid2008multiscale}. Alternative techniques \cite{kohler2023flow, arts2023two} have been developed that first train a generative model and extract a PMF from the model, yet have not been as widely used in practice.

Early coarse-grained ML potentials \cite{wang2019machine, husic2020coarse} demonstrated the ability to reproduce the free energy surface of chignolin after training on 200 $\mu$s of short simulations. Since then, the focus has been on generating larger, more diverse training sets to develop \textit{transferable} potentials. Majewski et al. \cite{majewski2023machine} aggregated short trajectories totaling 200--2000 $\mu$s for each of 12 fast-folding proteins; a single neural network trained on this data could reproduce the free energy landscapes and native structures of all 12 proteins. Charron et al. \cite{charron2025navigating} simulated 50 CATH domains for 2 $\mu$s each to develop the first fully transferable coarse-grained ML potential. This potential was demonstrated to reproduce the free energy landscapes and folding pathways of miniproteins (up to 73 AA) with low sequence similarity to training domains, the ligand-dependent folding of an IDP in the PUMA-MCL1 dimer (189 AA), and mutational ddGs of ubiquitin (76 AA). 

A key advantage of ML potentials is that, as direct substitutes for classical force fields, they can be combined with enhanced sampling methods and integrated into computational protocols that rely on simulations, such as transition path sampling and free energy calculations. On the other hand, while coarse-grained potentials can reduce the number of integration steps, neural network evaluations are strictly slower (oftentimes by orders of magnitude) than classical force fields per integration step. Thus, simulating long-timescale motions of moderate or large protein systems with coarse-grained ML potentials could still remain challenging.

\section{Generative models of ensembles}

Generative models are deep learning frameworks that parameterize and enable the sampling of high-dimensional, multimodal distributions. They have received widespread attention for numerous applications in biomolecular modeling, such as structure prediction, molecular docking, and protein design; see Yim et al. \cite{yim2024diffusion} for a broader review. In application to protein ensembles, generative models can be trained to sample molecular configurations conditioned on the identity of a molecular system, providing a new paradigm for sampling equilibrium distributions that eschews physical simulation. We refer to such models as \textit{ensemble emulators}. A defining property of these models is that they draw statistically independent samples with fixed computational cost, thus overcoming the curse of correlated samples that plagues molecular dynamics. Thus, they represent perhaps the most ambitious and end-to-end paradigm for AI-based sampling of protein ensembles.

\subsection{Transferable ensemble emulators}
The first ensemble emulators were trained on MD trajectories of individual proteins and could sample diverse, low-energy, physically plausible configurations of those systems \cite{degiacomi2019coupling,noe2019boltzmann}. However, the dependency on MD simulations for the system of interest limited the practical utility of such methods. Major strides in developing \textit{transferable} models were enabled by architectural elements from AlphaFold2 \cite{jumper2021highly}, which is manifestly transferable across protein sequences. Specifically, features extracted from MSAs by folding models can be used to condition a diffusion model to produce a distribution of structures \cite{jin2025p2dflow,fan2024accurate,zheng2024predicting,lewis2025scalable,wang2024protein}. Following this workflow, AlphaFlow \cite{jing2024alphafold} and UFConf \cite{fan2024accurate} could improve recall of PDB conformational states relative to AlphaFold2 on a curated set of conformational changes \cite{fan2024accurate} (up to 768 AA).

To increase and better assess sample diversity, an emerging body of work proposes to fine-tune and validate models on molecular dynamics trajectories after pre-training on the PDB. Several of these efforts make use of recent public MD simulation datasets such as ATLAS \cite{vander2024atlas} and mdCATH \cite{mirarchi2024mdcath}. Distributional Graphormer (DiG) \cite{zheng2024predicting} reported coverage of conformational states of the SARS-Cov-2 RBD observed in longer simulations. AlphaFlow \cite{jing2024alphafold} systematically assessed ensemble accuracy on 82 test proteins and reported recovery of root mean square fluctuation (RMSF) profiles, contact fluctuations, and solvent exposure events. Similar models, with variations in architecture, have been trained and validated on MD datasets \cite{jin2025p2dflow, wang2024protein, wolf2025learning,lu2024structure,janson2025deep}. To date, BioEmu \cite{lewis2025scalable} represents the largest effort towards developing a transferable protein ensemble emulator. Trained on AlphaFoldDB with a novel data augmentation procedure and fine-tuned on 200 ms of in-house MD trajectories, BioEmu reported recovery of alternative conformational states, local unfolding events, conformational flexibility in IDRs, and accurate prediction of folding free energies.

Post-AlphaFold2, several structure prediction models have been natively trained as generative models, including AlphaFold3 \cite{abramson2024accurate} and similar models \cite{boitreaud2024chai,wohlwend2024boltz}. These are also, in principle, capable of modeling protein ensembles, and blur the line between structure predictors and ensemble emulators. However, AlphaFold3 has been shown to underperform AlphaFlow and UFConf on a curated set of conformational changes \cite{fan2024accurate} and produces low sample diversity \cite{abramson2024accurate}. Boltz-2 \cite{passaro2025boltz} reports improved sample diversity by fine-tuning on MD data. Further evaluation is needed to fully articulate the relationship between general-purpose structure predictors and bespoke ensemble generators.

Separately from efforts building on AlphaFold2, transferable generative models that target conformational ensembles of short peptides \cite{abdin2023pepflow} and intrinsically disordered proteins (IDPs) \cite{janson2023direct,janson2024transferable,zhu2024precise,zhang2025deep} have been developed. Due to scant experimental data, the training of these models relies heavily on the collection of MD ensembles, often from IDP-tailored, coarse-grained force fields. These methods have shown success in recapitulating experimental observables such as radii of gyration and nuclear magnetic resonance (NMR) chemical shifts.

Despite rapid progress, the value of current ensemble emulators remains difficult to evaluate. There is a lack of consensus on common evaluation metrics between methods, and ground truth ensembles are difficult to acquire via either simulation or experiment. Evaluations based on PDB structures are appealing, but can be misleading due to incomplete coverage of protein states, absence of statistical weights, and ambiguity between induced conformational changes and inherent flexibility. On the other hand, the diversity and simulation length of public MD datasets remains limited (sub-$\mu$s trajectories for $\sim$1000 proteins), and methods tuned on such trajectories are unlikely to capture long-timescale events of biological interest. Conversely, methods benchmarked on recall of known conformational events (e.g., BioEmu \cite{lewis2025scalable}) may also predict spurious flexibility \cite{lewis2025scalable,sledzieski2025rocketshp,passaro2025boltz}. A major challenge is to disentangle structural diversity arising from model uncertainty about the conformational landscape versus true conformational flexibility. To date, no method can accurately reproduce free energy landscapes or statistical weights in a transferable manner. Further progress will require deliberation on how best to evaluate generated ensembles even when they fall short of this gold standard.

\subsection{Boltzmann reweighting}
A risk with generative models is the sampling of inaccurate ensembles without a means for detecting and correcting such errors. Boltzmann generators \cite{noe2019boltzmann} reduce this risk by parameterizing the generative model with normalizing flows, a class of architectures that provide model densities for generated samples. This approach enables reweighting of samples with physical potentials and asymptotically unbiased estimation of observables, even if the generative model is imperfect. Noe et al. \cite{noe2019boltzmann} demonstrated that such Boltzmann generators, trained on the long DESRES simulation of BPTI (58 AA) \cite{shaw2010atomic}, could reproduce distributional statistics to high fidelity and estimate temperature-dependent free energy differences between macrostates. As such, Boltzmann generators have long appeared highly attractive for integrating generative AI with principled physics-based sampling.

However, the development of a transferable Boltzmann generator for proteins has proven challenging. Compared to diffusion models, normalizing flows are more constrained architectures and are harder to generalize across protein systems. In fact, the BPTI Boltzmann generator of Noe et al. \cite{noe2019boltzmann} relied on bespoke internal coordinate transformations requiring advance knowledge of its conformational landscape, manifestly precluding transferability. More expressive \textit{continuous} normalizing flows have enabled the development of transferable Boltzmann generators for dipeptides \cite{klein2024transferable}; but the cost of likelihood evaluation grows with the squared dimensionality of the system, making scaling to large proteins speculative. More recently, transferable transformer-based Boltzmann generators capable of sampling ensembles of short peptides (up to 10 AA) have been reported \cite{tan2025amortized}. That said, all known methods suffer from low effective sample size for systems larger than alanine dipeptide. The development of expressive, fast-likelihood, transferable architectures to enable self-correcting ensemble emulators remains an active area of research.

\subsection{Training without simulations}
Generative models are typically trained with data from the target distribution; however, obtaining converged ensembles for a large number of systems is manifestly challenging. As such, a longstanding goal is to develop protocols for training generative models \textit{with access to the target energy alone}. This can be thought of as directly distilling a physical potential into the generative model, without simulated ensembles as an intermediate. Noe et al. \cite{noe2019boltzmann} laid out the groundwork for this approach with normalizing flows, proposing to use the model likelihood to directly minimize the reverse Kullback-Leibler divergence to the Boltzmann distribution. Following this approach, FAB \cite{midgley2023flow} and TA-BG \cite{schopmans2025temperature} have demonstrated energy-based learning on alanine di-, tri-, and hexapeptides; yet, these approaches operate on internal coordinates, precluding transferability. 

More recently, attention has turned to frameworks for energy-based training of diffusion models, often called \textit{diffusion samplers} \cite{vargas2023denoising,akhound2024iterated,havens2025adjoint}. PITA \cite{akhound2025progressive} demonstrated pure energy-based training on alanine di- and tri-peptide in \textit{Cartesian} coordinates. Transferable methods for similarly-sized systems may be developed in the near future, but the computational efficiency of diffusion sampler training remains a concern. Replica exchange dynamics have been shown to explore a simple Gaussian mixture model more efficiently than diffusion samplers, even without accounting for the cost of neural network evaluations \cite{he2025no}. A potentially more scalable approach would be to use diffusion sampler training in combination with data-based training \cite{abdin2023pepflow,zheng2024predicting}. Rigorous and creative benchmarking efforts are needed to establish in what regimes, if any, energy-based training is most effective.

\section{MSA perturbation methods}

AlphaFold and similar methods rely on multiple sequence alignments (MSAs) to predict protein structures; yet different MSAs may lead to different structures for the same sequence. Strategies for perturbing these MSAs have emerged as a surprisingly effective means for predicting protein ensembles or alternate states with AlphaFold2 off-the-shelf. Del Alamo et al. \cite{del2022sampling} first reported that random subsampling of the MSA reveals open and closed states in membrane proteins. Using similar techniques, da Silva et al. \cite{monteiro2024high} were able to predict mutational effects on the relative population of active and inactive states of Abl kinase. CF-Random \cite{lee2025large} showed successful predictions of diverse motions, including fold switching, rigid body motions, and local conformational changes. Taking an alternative approach, SPEACH\_AF \cite{stein2022speach_af} and AFSample2 \cite{kalakoti2025afsample2} systematically mask or mutate columns of the MSA to sample the conformational landscape of membrane proteins. Building on the hypothesis that MSAs reveal competing coevolutionary signals, AFCluster \cite{wayment2024predicting} predicts one structure per sequence cluster within the MSA. This pipeline was able to predict alternative folds of fold-switching proteins and led to the discovery (and validation) of a previously unknown fold-switcher.

Despite the efficacy and popularity of MSA perturbation methods, we note that they arose as fortuitous workarounds to the limitations of structure predictors rather than a design principle for modeling ensembles. Although \textit{post-hoc} hypotheses for their efficacy have been proposed \cite{wayment2024predicting,lee2025large}, the training of structure prediction models \textit{penalizes} sensitivity to different MSAs for the same sequence. Due to this misalignment, it is unclear how to improve methods via training on additional ensemble data. Furthermore, the input sensitivity may be reflective of model uncertainty and is not always necessarily correlated with physical flexibility. Perturbations can degrade prediction quality and introduce spurious flexibility not reflective of the physiological ensemble \cite{jing2024alphafold,lewis2025scalable}. This effect is reflected in the poorer performance relative to generative models on benchmarks of smaller conformational changes \cite{fan2024accurate}. On the other hand, predicted structures inherit AlphaFold2's bias towards low-energy structures and do not reflect fluctuations at physiological temperatures \cite{jing2024alphafold}. 

Nevertheless, MSA perturbation methods are the most mature and practically useful family of methods discussed in this review. They have successfully modeled a diverse array of large, biologically relevant motions, have led to experimentally validated discoveries, inherit the widespread acceptance of AlphaFold2, and are easy to use. As such, they have become adopted as the \textit{de facto} standard method of generating protein ensembles in many studies.

\section{Local transition models}
An emerging set of approaches seeks to evolve protein configurations \textit{locally}, defining a Markov chain that explores the conformational landscape. This is an easier problem than directly generating independent samples, as the model needs only to reason about a small neighborhood of conformations rather than the global landscape. Str2Str \cite{lu2024strstr} (for proteins) and JAMUN \cite{daigavane2024jamun} (for peptides) define this transition via partial noising and denoising using a transferable diffusion model; conceptually this can be thought of as simulating with a smoothened energy function. Although trained on PDB data, Str2Str \cite{lu2024strstr} could recapitulate some characteristics of fast-folding simulations from DESRES \cite{lindorff2011fast}. Another set of approaches, sometimes called \emph{surrogate simulators}, target the transitions observed in molecular dynamics simulations. Although such simulators have been previously proposed \cite{tsai2020learning,sidky2020molecular}, recent approaches are characterized by stochastic transitions and tranferable architectures. TimeWarp \cite{klein2023timewarp} and MDGen \cite{jing2024generative} are trained to match 100-ns transitions of short peptides and reproduced free energy landscapes of unseen peptides. ITO \cite{schreiner2023implicit}, F3low \cite{li2024f}, and EquiJump \cite{costa2024equijump} are trained and validated on the 12 fast-folding protein simulations from DESRES \cite{lindorff2011fast}, accurately capturing their free energy landscapes. A fully transferable local generative model for protein systems is not yet available, but should be relatively straightforward to train using the same datasets \cite{vander2024atlas,mirarchi2024mdcath} used to train ensemble emulators.

Since the samples are correlated, statistical convergence and efficient sampling of long-timescale events return as important concerns. If matching transition operators, one can strategically choose between very long intervals, approaching independent sampling; and very short intervals, minimizing the difficulty but also sampling efficiency. Another key challenge is to ensure that the stationary distribution indeed approximates the desired protein ensemble. Since sampled conformations are fed back into the model, errors could accumulate towards hallucinated or pathological states. TimeWarp \cite{klein2023timewarp} attempted to rigorously avoid this possibility by using a normalizing flow architecture and its exact likelihoods for Metropolis-Hastings rejections, functioning as an MCMC proposal. However, the model struggled to produce useful acceptance rates. Although ITO \cite{schreiner2023implicit} and EquiJump \cite{costa2024equijump} demonstrated stable rollouts and good reconstruction of free energy landscapes without rejection steps, transferable models evaluated on unseen proteins may struggle more.

\section{ML collective variables}

Instead of entirely circumventing simulation, ML techniques have also been developed to augment established enhanced sampling algorithms. We defer to recent comprehensive reviews  \cite{mehdi2024enhanced,gokdemir2025machine} for a systematic exposition of ML-based enhanced sampling. For proteins, the most mature approaches are those that use neural networks to parameterize \textit{collective variables} for algorithms such as metadynamics. Several theoretical frameworks exist to learn optimal collective variables (CVs) from simulation data of individual systems \cite{mardt2018vampnets,wang2021state}. To make such frameworks practically useful, workflows involving iterative learning of CVs and enhanced simulations have been developed \cite{bonati2021deep,mehdi2022accelerating}. However, to our knowledge, such workflows have yet to be demonstrated on systems larger than Trp-cage (18--20 AA) when starting from only a single structure. For purposes of developing transferable methods for protein ensembles, the promise of ML-based CVs lies more in the ability to leverage \textit{prior knowledge} about protein ensembles in order to obtain a CV for the target ensemble.

AF2RAVE \cite{vani2023alphafold2} represents a pioneering attempt to leverage such prior knowledge. In this protocol, diverse conformations predicted by MSA perturbation methods are used to seed the initial round of simulations for building the ML collective variable. This protocol has been demonstrated on diverse proteins over 200 AA in several biologically relevant contexts \cite{gu2025hierarchical,vani2023exploring}. We distinguish, however, between the transferable MSA perturbation method and the learned collective variable, which is \textit{not} transferable. The development of a \textit{bona fide} transferable neural network that directly parameterizes the collective variables, trained and refined across systems, could be highly attractive.

Unlike most other methods discussed in this review, machine-learned CVs with enhanced sampling algorithms can guarantee correct thermodynamic weights under the Boltzmann distribution. Moreover, these techniques can be more easily adapted to contexts unseen during training, such as modified temperatures or intermolecular interactions. Although faster than unbiased simulation, enhanced sampling methods still incur significant cost; for example, AF2RAVE required 2 $\mu$s of aggregate  simulation to explore the conformational landscape of a 76 AA protein \cite{vani2023alphafold2}. As such, these methods may not be well suited for high-throughput ensemble prediction.

\section{Predicting ensemble descriptors}

In lieu of generating full protein structures to describe conformational diversity, models have been proposed that predict \textit{descriptors} of protein flexibility or dynamics, such as principle motions, RMSF profiles, and pairwise correlations from protein sequence or structure. Although they do not provide ensembles \textit{per se} and are strictly less informative about the conformational landscape, these descriptors can already be useful for biological analysis. For example, pairwise correlations between residues can be sufficient to study allosteric networks, i.e., patterns of motion involved in allosteric conformational changes \cite{bowerman2016detecting}. These predictors, usually fine-tuned from protein language models and trained on MD data, are extremely fast due to the simpler output space and requiring only a single evaluation. Nonetheless, they may obtain higher accuracy on directly supervised quantities compared to extracting descriptors from an imperfectly predicted ensemble. 

Current predictors---each with a different set of output descriptors---include RocketSHP \cite{sledzieski2025rocketshp} and DynaProt \cite{bafna2025learning}, which require a static structure input; and SeaMoon \cite{lombard2025seamoon}, SeqDance \cite{hou2024seqdance}, and Dyna-1 \cite{wayment2025learning}, which do not. DynaProt \cite{bafna2025learning} and SeaMoon \cite{lombard2025seamoon} predict anisotropic descriptors of motion---akin to normal mode analysis---enabling imputation of a crude structural ensemble or alternative states. While most methods are supervised on training targets derived from simulated ensembles, Dyna-1 \cite{wayment2025learning} is notable for being primarily trained on signatures of long-timescale motions in \textit{experimental} NMR data.

\begin{figure}
    \centering
    \includegraphics[width=0.9\linewidth]{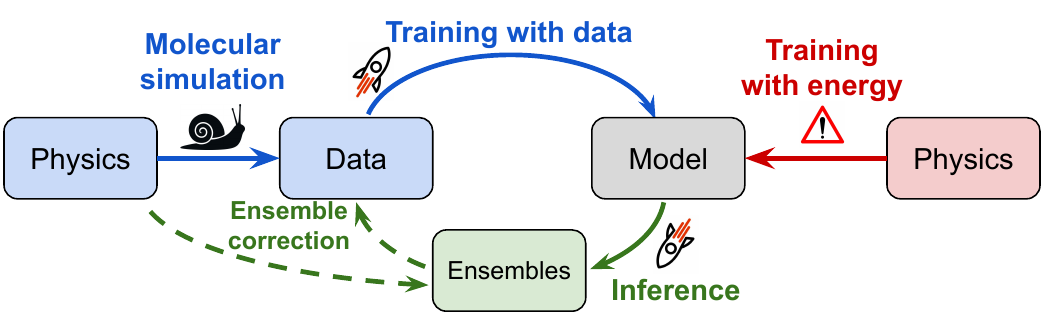}
    \caption{``Closing the loop" as a strategy for overcoming challenges with scaling current methods beyond available datasets. In the canonical training workflow (blue), physics-based molecular simulation generates training data, but the cost of simulation presents a major bottleneck. Techniques that train models with direct access to the energy (red) have failed to scale beyond toy systems. We advocate for the development of ``ensemble correction" algorithms that use physics to improve the quality of ensembles generated from the model. For a sufficiently well-trained model, this can generate new training data faster than molecular simulation, allowing for accelerating model self-improvement (green).}
    \label{fig:loop}
\end{figure}

\section{Outlook}

A major factor for the future improvement of nearly all methods is the availability of training data. Public simulation datasets are growing \cite{vander2024atlas,mirarchi2024mdcath} but will remain insufficient in the foreseeable future, barring a several orders-of-magnitude increase in resources dedicated to MD simulation. Meanwhile, methods for learning from energy alone are not yet effective on systems of realistic size. To overcome these barriers, we advocate for ``closing the loop" between data generation, training, and inference (Figure~\ref{fig:loop}). Namely, \textit{trained models should be leveraged in combination with simulation and/or physical potentials to collect data for the next phase of models}. In principle, any framework for using physical potentials to improve accuracy at inference time can serve as training signal, as explored in many energy-based training frameworks. However, we  particularly advocate for \textit{post-hoc} corrections to generated ensembles, which may be more versatile and scalable. For example, enhanced sampling with prior knowledge derived from approximate ensembles, as illustrated in AF2RAVE \cite{vani2023alphafold2}, already provides a concrete instantiation of such a procedure. Alternatively, predicted ensembles descriptors could potentially be used to define transferable collective variables to bias the next phase of simulations. The design space of these ``ensemble correction" algorithms is likely vast; we contend that they will be essential for the eventual development of accurate, scalable, and transferable methods.

Any techniques that rely on simulated data or physical potentials will also inherit the inaccuracies of the underlying force field, which can be significant. \textit{Ab initio} ML potentials with improved accuracy are under development \cite{wang2024ab}, but will be even more expensive than classical simulations. Thus, it is imperative for future methods to remain grounded in experimental data, including from techniques such as SAXS, smFRET, and HDX-MS, which do not provide structural configurations. Although inference-time techniques for guiding structural models with experimental constraints have been developed \cite{fadini2025alphafold}, new algorithmic frameworks will be needed to \textit{train} transferable structural models using non-structural experimental labels. The property-prediction fine-tuning algorithm in BioEmu \cite{lewis2025scalable} represents an encouraging first step.

Finally, transferability across physicochemical conditions, i.e., temperature, pH, or ionic strength, is often assumed in all-atom MD simulation, whereas AI-based methods must be deliberately trained across conditions to possess such transferability. However, the generation of training data at all possible conditions of interest is impractical. To increase the versatility of AI-based methods, frameworks that permit perturbation of the target distribution to conditions unseen at training time are likely needed. This will require models to interface with the force field at inference time, rather than only at training time or for data generation, as most current methods do.

\section*{Acknowledgements}
We thank Mihir Bafna, Samuel Sledzieski, Peter Holderreith, Soojung Yang, Juno Nam, Ezra Erives, Yuanqi Du, and Allan dos Santos Costa for helpful feedback and discussions. This work was supported by the National Institute of General Medical Sciences of the National Institutes of Health under award 1R35GM141861 (to B.B.), a Department of Energy Computational
Science Graduate Fellowship under Award Number DESC0022158 (to B.J.), the Machine Learning for Pharmaceutical Discovery and Synthesis (MLPDS) consortium, the DTRA Discovery of Medical Countermeasures Against New and Emerging (DOMANE) threats program, the NSF Expeditions grant (award 1918839) Understanding the World Through Code, and the DSO Singapore grant on next generation techniques for protein ligand binding.
\newpage

\bibliography{references}

\end{document}